\documentclass[conference,a4paper]{IEEEtran}
\usepackage{multicol}
\usepackage{algorithm}
\usepackage{algpseudocode}
\usepackage{breqn}
\usepackage{color}
\usepackage{soul}
\usepackage{amsmath}
\usepackage{graphics}
\usepackage{setspace}
\usepackage{cite}
\usepackage{latexsym}
\usepackage{float}
\usepackage{epsfig}
\usepackage{multirow}
\usepackage[table,xcdraw]{xcolor}
\usepackage{cite,cases,url}
\usepackage{amssymb}
\usepackage{graphicx}
\usepackage{epstopdf}
\usepackage{balance}
\usepackage{enumerate}
\usepackage{array}
\usepackage{algorithm}
\usepackage{enumitem}

\usepackage[normalem]{ulem}
\useunder{\uline}{\ul}{}
\DeclareUnicodeCharacter{2212}{-}
\newcolumntype{L}{>{\centering\arraybackslash}m{5cm}}
\newcolumntype{K}{>{\centering\arraybackslash}m{6cm}}
\newcolumntype{P}{>{\centering\arraybackslash}m{2.3cm}}
\newcolumntype{M}{>{\raggedright\arraybackslash}m{2cm}}
\newcolumntype{N}{>{\raggedright\arraybackslash}m{2.5cm}}

\begin{document}

\pagenumbering{gobble}

\title{Analysis of Reinforcement Learning Schemes for Trajectory Optimization of an Aerial Radio Unit
}

\author{
\IEEEauthorblockN{Hossein Mohammadi$^{\dag}$, Vuk Marojevic$^{\dag}$, and Bodong Shang$^{\star}$
}
\normalsize\IEEEauthorblockA{$^{\dag}$Department of Electrical and Computer Engineering,  Mississippi State University,
Mississippi State, MS, USA\\
$^{\star}$Carnegie Mellon University, Pittsburgh, USA}
$\{$hm1125$|$vuk.marojevic$\}$@msstate.edu. bdshang@cmu.edu
\vspace{-3mm}
}

\maketitle

\begin{abstract}
This paper introduces  the deployment of unmanned aerial vehicles (UAVs) as lightweight wireless access points that leverage the fixed infrastructure in the context of the emerging open radio access network (O-RAN). More precisely, 
we propose an aerial radio unit that dynamically serves an under served area and connects to the distributed unit via a wireless fronthaul between the UAV and the closest tower.  In this paper we analyze the UAV trajectory in terms of artificial intelligence (AI) when it serves both UEs and central units (CUs) at the same time in multi input multi output (MIMO) fading channel. We first demonstrate the nonconvexity of the problem of maximizing the overall network throughput based on UAV location, and then we use two different machine learning approaches to solve it. We first assume that the environment is a gridworld and then let the UAV explore the environment by flying from point A to point B, using both the offline Q-learning and the online SARSA algorithm and the achieved path-loss as the reward. With the intention of maximizing the average payoff, the trajectory in the second scenario is described as a Markov decision process (MDP). According to simulations, MDP produces better results in a smaller setting and in less time. In contrast, SARSA performs better in larger environments at the expense of a longer flight duration.
\end{abstract}

\IEEEpeerreviewmaketitle

\section{Introduction}
\label{sec:intro}
\indent Drones, also referred to as unmanned aerial vehicles (UAV), have gained significant interest in recent years for a variety of uses, including improving coverage for large cities and delivering freight\cite{01wu2018joint}. According to Morgan Stanley's estimation, the worldwide market for urban air mobility (UAM), including drone deliveries and air taxis, may account for up to $11\%$ of the predicted GDP for the entire world \cite{03jonas2021evtol}. Aerial Base Stations (ABSs) will provide connectivity to ground users in future 6G networks. 
Moreover, according to shorter-term forecasts, the recreational and commercial UAV fleets are expected to reach up to 3 million by 2023, and the drone services market is expected to be worth more than 63.6 billion dollars by 2025. \cite{02colpaert2022drone}.\\
\indent The performance of UAVs in a cellular network has been 
investigated using stochastic geometry \cite{04azari2019cellular}. Due to interference issues and high handover rates, as well as other simulation-based research, it has been demonstrated that present LTE networks do not provide adequate coverage above building height \cite{05colpaert20203d}. However, as we can see in some of the studies in this field for reducing the interference for femtocells, the authors in \cite{09saquib2012interference} present an overview of the many cutting-edge methods for managing resources and interference in orthogonal frequency division multiple access (OFDMA)-based femtocell networks where it includes a qualitative comparison of the various methods. For this reason, unresolved issues with OFDMA femtocell network interference management scheme design are highlighted in \cite{09saquib2012interference}. In terms of coverage and handover rates, a number of measurement studies produced satisfactory results. However, as opposed to an urban setting, they were carried out in rural areas where the BS density is significantly lower \cite{06gharib2021exhaustive}.\\
\indent Although there are opportunities using scalable learning techniques for data-driven wireless networks, these techniques have their own problems. First, in order to reduce the load on central nodes and decrease latency, computations will likely be shifted to edge devices in the future. Therefore, in order to work within the limitations of on-device computation, storage, and battery capacities, low-complexity learning models must be developed. Second, enhancing the interpretability of data-driven models by revealing their "black box" is undoubtedly a critical area for research \cite{17xu2020scalable}. One suggested solution is that of a self-organizing network (SON) to enable the automatic deployment of cellular networks. 
The deployment difficulties include, among others, data imbalance, data insufficiency, cost insensitivity, and non-real-time reaction \cite{07zhang2022data}. However, by increasing the global network and the volume of data transferred among devices; these approaches are not practical anymore, in particular in B5G and 6G.\\
\indent Network entities in UAV networks must decide locally to optimize network performance in the face of a network environment that is unknown. When the state and action spaces are small, reinforcement learning has effectively been employed to enable network entities to achieve the best policy for decisions or actions given their states. Reinforcement learning, however, may not be able to establish the ideal policy in a reasonable amount of time in complicated and large-scale networks since the state and action spaces are typically large. In order to address this, deep RL (DRL), a blend of reinforcement learning and deep learning, was created \cite{08luong2019applications}.\\
\indent 
Data transmission through UAVs has been studied in \cite{18dai2022joint}, which provides a data delivery scheme to address the pricing issue for UAVs and a data delivery strategy for users to increase the effectiveness of data transmission with the aim of maximizing the functionality of both UAVs and users. 
This paper introduces  the deployment of UAVs as lightweight wireless access points that leverage the fixed infrastructure in the context of the emerging open radio access network (O-RAN). More precisely, 
we propose an aerial radio unit that dynamically serves an under served area and connects to the distributed unit via a wireless fronthaul between the UAV and the closest tower. \\


\vspace{-5mm}
\section{Related Work}
\label{sec:related}
\indent This section's literature review 
assesses the feasibility of UAV data transmission and discusses the benefits and drawbacks of 
artificial intelligence (AI) 
for UAV trajectory optimization.\\
\indent The authors of \cite{01wu2018joint} maximize the lowest throughput over all ground users in the downlink connection by concurrently optimizing multiuser communication scheduling and association with the UAV trajectory and power control in order to achieve equitable performance among users. The formulated problem is a 
mixed integer nonconvex optimization problem. As a result, they suggest an effective iterative solution for solving it using consecutive convex optimization and block coordinate descent. In each iteration 
the user scheduling and association, UAV trajectory, and transmission power are alternately optimized. 
\\
\indent In the context of full-duplex multi-UAV networks, \cite{10dai2022multi} investigates the 
decoupling of the uplink (UL)-downlink (DL) association and trajectory design challenge. The goal of the joint optimization task is maximizing the sum-rate of the UEs in both UL and DL. A robust partially observable Markov decision process (POMDP) model is presented to define the model uncertainty because the problem is non-convex with complex states and a single UAV cannot be aware of the reward functions of other UAVs. It is suggested to use a multi-agent DRL (MADRL) approach that allows each UAV to choose its policy in a distributed fashion.\\
\indent Several earlier studies looked on the user association issue for UAV-aided communications like \cite{12xi2019joint}–\cite{14liu2019mmwave}. For instance, 
\cite{11el2019distributed} investigates how to operate multi-UAV networks where UAVs can associate with terrestrial UEs and periodically update their locations with the aim of maximizing the accumulated DL rate. A joint optimization problem of UE-UAV link and UAV deployment 
is studied in \cite{12xi2019joint}. Reference \cite{14liu2019mmwave}  investigated how to link a terminal to a UAV in the DL and a ground station in the UL for a mmWave UAV network.\\
\indent Another approach is using beamforming or directional antennas. According to \cite{15feriani2022multiobjective} each base station has directional antennas with a two-dimensional radiation pattern, while the UEs feature omni-directional antennas with a 0 dBi antenna gain. In fact, the authors present 
a load balancing multi-objective reinforcement learning (MORL) framework. They provide a method based on meta-RL and develop a general policy that can rapidly adjust to new trade-offs between the objectives.\\
\indent Reference \cite{16noh2022ici} suggests designs for transmit beamformers and receive beamformers for a UAV network made up of a target, a UAV joint communication and radar (JCR) base station, and numerous user devices. The UAV broadcasts orthogonal frequency division multiplexing (OFDM) signals with transmit and receive beamforming for simultaneous MIMO radar and MU-MIMO communication tasks.\\

\vspace{-5mm}
\subsection{AI Models}
\indent 
Machine learning and deep learning, two subcategories of AI, have advanced to the point where they will support fifth generation (5G) and Beyond 5G (B5G) wireless networks. Additionally, a new era of wireless communications with full AI support is anticipated 
between 2027 and 2030 in order to meet the growing need for wireless connectivity. 
AI can help improve conventional schedulers and congestion control systems as well as reduce 
packet losses in vehicle to everything (V2X) connections, among others \cite{24mohammadi2022ai}.\\
\indent 
A variety of algorithms and approaches are used in deep learning to provide high-level representations of data. Deep learning's primary objective is to eliminate the need for manual data structure definition through automatic data-driven learning. Its name is an allusion to the fact that any neural network with two or more hidden layers is commonly referred to as a DNN. Artificial neural networks (ANN) are the basis for the majority of deep learning models. The ANN is a type of computational nonlinear model inspired by the neural organization of the brain that can be trained to carry out tasks including classification and prediction. Artificial neurons make up an ANN, which is divided into three interconnected layers called input, hidden, and output layers \cite{mohammadi2021artificial}.\\
\indent Over the past 20 years, the development of AI has been significantly impacted by RL. An agent can periodically make decisions, evaluate the outcomes, and then autonomously modify its approach to obtain the best possible policy through its interactions with the environment. 
The agent first observes its existing condition, then acts, reaping both its immediate reward and its new state. The agent's policy is adjusted based on the observed information, such as the immediate reward and the new state, and this process is repeated until the agent's policy approaches the optimal policy. DRL 
considerably speeds up the learning, especially for problems with large state and action spaces. As a result, DRL enables network controllers or IoT gateways to dynamically govern user association, spectrum access, and transmit power for a sizable number of IoT devices and mobile users in large-scale networks, such as IoT systems with thousands of devices \cite{08luong2019applications}.\\

\vspace{-5mm}
\section{System Model}
\label{sec:system}
The data rate 
and the signal-to-noise-plus interference ratio are defined as follows:
\begin{equation}
    R = log_2(1 + SINR_t),
\end{equation}
\begin{equation}
    SINR_t = \frac{P_{received power,t}}{\sigma^2 + \sum_k I_k},
\end{equation}
where $\sigma$ is the noise power and $I_k$ is the inter symbol interference back scattered from the environment.\\
For the received power we need to estimate the path loss:
\begin{equation}
    \begin{split}
        PL(dB) = 20log(\frac{4\pi f_cd}{c}) +&Prob(LoS) \eta_{LoS} +\\
                                         &Prob(NLoS)\eta_{NLoS}.
    \end{split}
    \label{eq:pl}
\end{equation}
The probability of line of sight (LoS) can be formulated as
\begin{equation}
    Prob(LoS)=\frac{1}{1+\mu_{1,2} exp(-\psi_{1,2}(\frac{180}{\pi})\theta_{i,j} - \mu_{1,2})},
\end{equation}
where $\mu_{1,2}$ and $\psi_{1,2}$ are the environment factors for the uplink and downlink channels and
\begin{equation}
    \theta_{i,j} = arctan(\frac{h_{i,j}}{l_{i,j}})
\end{equation}
is the angle with $h$ being the UAV altitude and $l$ the horizontal distance to the CU/DU. Moreover, $\eta_{LoS}$and $\eta_{NLoS}$ are additional losses due to the free space propagation \cite{25ghanavi2018efficient}. \\

\indent Moreover, in the uplink from UE to the UAV, the UAV adds nonlinearity to the received signal then retransmit it to the CU/DU. However, in the CU/DU in order to eliminate the interference and phase shift in the received signal we apply the model defined in \cite{22mohammadi2022self}. 
The MIMO channel model between the UE and the UAV and between the UAV and the CU/DU is obtained from \cite{23heath2016overview}. Consequently, the signal in the UAV receiver and CU is as follows:\\
\begin{equation}
    \begin{split}
        Y_{UL} &= XH_{UL} + \frac{3}{2}\beta|X|^2\\
        Y_{DL} &= Y_{UL}H_{DL} + \eta_{noise}.
    \end{split}
    \label{eq:channelmodel}
\end{equation}
The second term in $Y_{UL}$ is the receiver nonlinearity applied to the received signal at the UAV \cite{21dsouza2020symbol} and 
$\eta_{noise}$ is the additive white Gaussian noise of zero mean and $\sigma_{noise}$ variance. \\

\section{Problem Formulation}
\label{sec:problem}
If we assume that we have $I$ users and $J$ central and distributed units (CU+DU) with one UAV, in order to maximize the 
network throughput we need to maximize the following equation subject to certain criteria:

\begin{equation}
    \begin{split}
        min \sum_t max & \sum_{i,j} R_{i,t} + V_{j,t} R_{j,t}\\
        &s.t.\\
        &h_{min,t}<h_t<h_{max,t}\\
      &x_{min,t}<x_t<x_{max,t}\\
      &y_{min,t}<y_t<y_{max,t}\\
      &\sum_{j,t} V_{j,t}R_{j,t} > \sum_{i,t}R_{i,t}\\
      &\sum_{j,t} V_{j,t} = 1; V_{j,t} \in \{0,1\}\\
    \end{split}
    \label{eq:optmodel}
\end{equation}


$R_{i,t}$ is the data rate between the $i$th user and the UAV at time $t$ and $R_{j,t}$ is the data rate between the UAV and the CU/DU at time $t$. Parameters $h$, $x$, and $y$ determine the location of the UAV, $V_j$ is the connection of $j$th CU+DU to the UAV. Since in our case we have only one UAV, there is no constraint on the UAV connection to the CU+DU. Moreover, the capacity of the link between the UAV and the CU+DU should be larger than that of the uplink from nodes to UAV. This assumption stems from the fact that several nodes might send data to the UAV.\\

\vspace{-5mm}
\section{Proposed Approach} 
\label{sec:contribution}

\begin{figure}
    \centering
    \includegraphics[scale = 0.2]{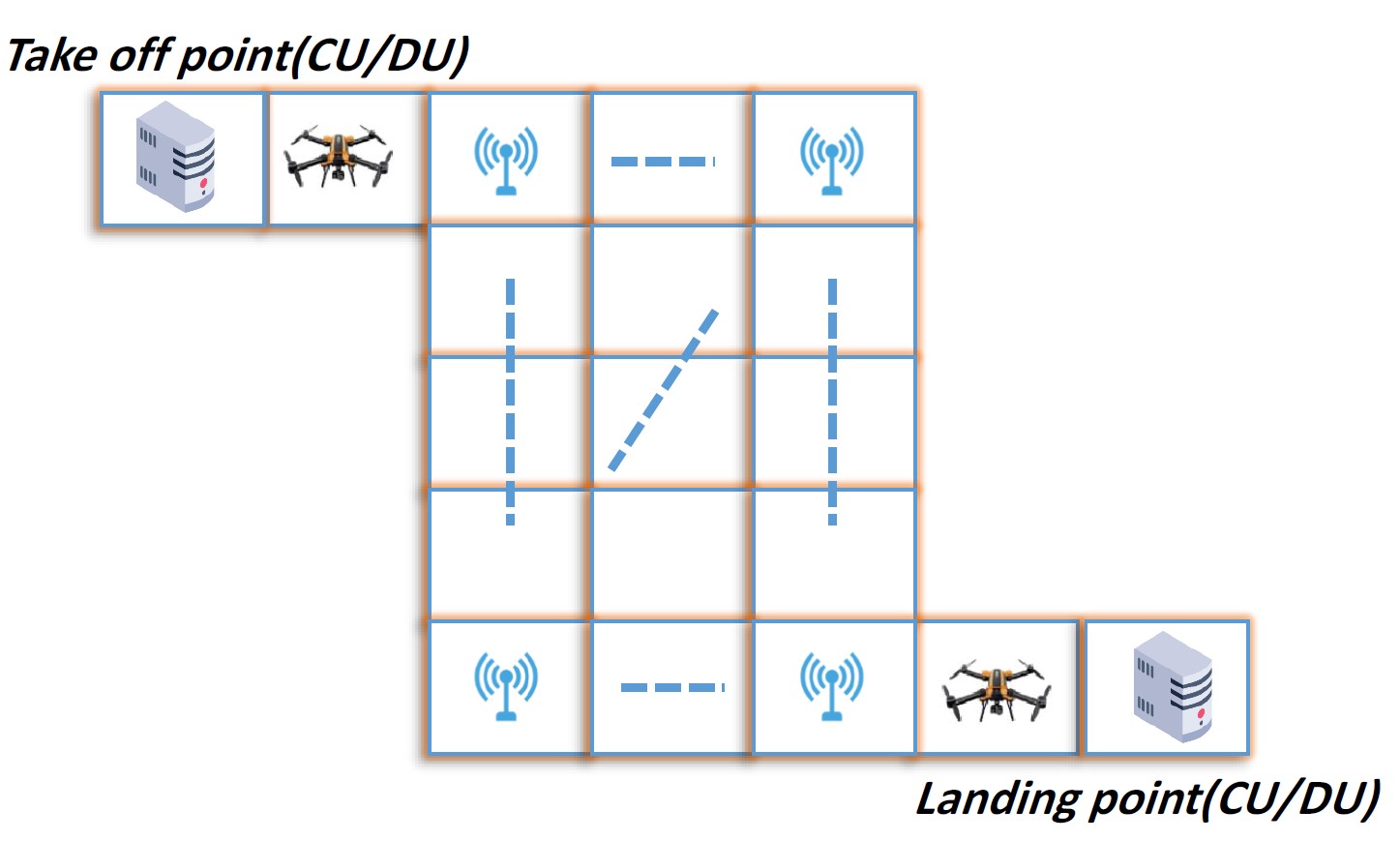}
    \caption{One scenario in RL is defining the initial and termination point for the UAV and let it move around to collect data from the UEs. Based compatible environment is the grid world}
    \label{fig:gridworld}
\end{figure}

\begin{figure}
    \centering
    \includegraphics[scale=0.25]{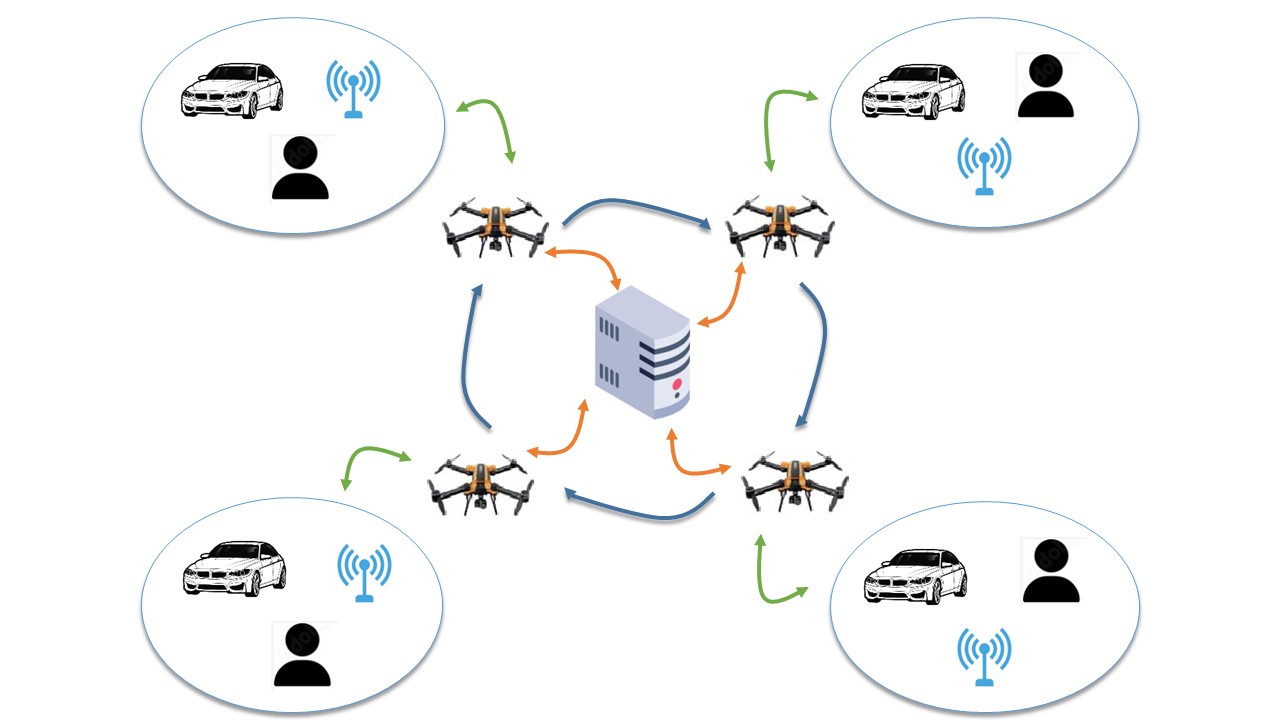}
    \caption{One UAV is moving in a pre-defined area to collect the data and send them to the CU}
    \label{fig:UAV movement}
    \vspace{-5mm}
\end{figure}


\indent As shown in Figs. "fig:gridworld" and "mdp,, we are required to examine the UAV trajectory under two distinct conditions. In one scenario, the UAV can choose from four possible actions, but passing through a previous state would result in a negative reward. In the second scenario, which is based on a Markov decision process (MDP) with two distinct UAV action options—"up" and "down"—the UAV must skip passing through some states in order to reach the terminal state with the maximum reward. For both cases, minimum steps to get the terminal states should be considered to minimize the UAV power consumption. Moreover, we investigate the UAV trajectory for each situation using two algorithms: Q-learning and SARSA. The first is for the scenario where the UAV operates offline, and the second is for the scenario where the UAV maximizes the average reward while operating online.

\begin{figure}
    \centering
    \vspace{-2mm}
    \includegraphics[scale=0.3 ] {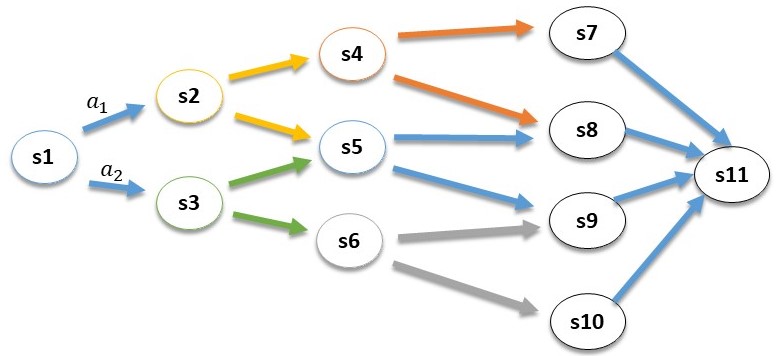}
    \vspace{-2mm}
    \caption{In the second scenario we are assuming the UAV movement follows the MDP algorithm in which it passes through the states with lower received power}
    \vspace{-8mm}
    \label{fig:mdp}
\end{figure}



\subsubsection{Q-Learning}
\indent In an MDP, our goal is to identify the best course of action for the agent in order to maximize the anticipated long-term reward function of the system. We first develop a value function $\nu^\pi : S \to \mathbb{R}$ that depicts the predicted value attained by adhering to state-specific legislation in which each state is in the set $s\in S$. With an unlimited horizon and discounted MDP, the value function 
can be defined as follows, 
which quantifies the goodness of the policy:
\begin{equation}
\begin{split}
    \nu^\pi (s) &= E_\pi \left [\sum_{t=0}^{\infty}\gamma r_t(s_t,a_t)|s_0=s \right]\\
    & = E_\pi \left[ r_t(s_t,a_t) + \gamma \nu^\pi(s_t+1)|s_0=s \right],
\end{split}
\label{eq;eq4}
\end{equation}
where $r_t(s_t,a_t)$ is the achieved reward at iteration $t$ when the agent is at state $s_t$ and takes action $a_t$. Parameter $\gamma\in [0,1]$ is the discount factor. In comparison to the current reward, the discount factor defines how important future rewards are. If the agent is "myopic," ($\gamma=0$) it solely thinks about how to maximize its current benefit, or short-term reward. In contrast, the agent will aim for a long-term larger reward if $\gamma$ is close to one. The key here is how much attention the UAV should give to potential rewards. Our suggested approach is to determine this value when, given the first observation of the environment, the assessment of the discounted long-term reward at the start of the episode is close to the average reward, which is defined as
\begin{equation}
    \frac{|Q_0 - AR|}{|Q_0 + AR|} \leq \Delta,
    \label{eq:DFThreshold}
\end{equation}
where $\Delta$ is the predefined threshold to select the best discount factor.\\
\indent The best action at each state can be discovered by using the best value function 
\begin{equation}
    \nu ^* (s) = max_{a_t}\Bigl\{E_\pi \left[r_t(s_t,a_t) + \gamma \nu^\pi (s_{t+1})\right] \Bigr\}.
    \label{eq:eq5}
\end{equation}

Now by defining 
\begin{equation}
    Q^*(s,a) \triangleq r_t(s_t,a_t) + \gamma E_\pi \left[\nu^\pi(s_t+1) \right],
    \label{eq:eq6}
\end{equation}
if $\nu^*=max_a{Q^*(s,a)}$ is the best Q-function for all state action pairings, then $\nu^*$ is the best value function. Currently, the objective is to determine the best Q function $Q^*(s,a)$ values for all state-action pairs, which can be accomplished through an iterative procedure:
\begin{multline}
    Q_{t+1}(s,a) = Q_t(s,a) + \alpha_t\Bigl[r_t(s,a) +\\
    \gamma max_{a^\prime}Q_t(s,a^\prime) - Q_t(s,a)\Bigr].
    \label{eq:eq7}
\end{multline}
The learning rate $\alpha_t \in [0,1]$ is utilized in (\ref{eq:eq7}) to assess how new knowledge will affect the current Q value. The learning rate can be set to be constant or it can change dynamically as the learner advances through the material. 
The Q-learning algorithm operates offline even though it can determine the agent's best course of action without knowing anything about the surrounding environment. 


\subsubsection{SARSA}
\indent Online Q-learning which is named the State-Action-Reward-State-Action (SARSA) algorithm is one of the well-known approaches in RL. The SARSA algorithm, as opposed to the Q-learning algorithm, is an online method that enables the agent to select the best course of action at each time step in real-time without having to wait for the algorithm to converge. In the Q-learning process, the policy is changed using an off-policy strategy that considers the behaviors with the highest potential reward. As an on-policy method, the SARSA algorithm engages with the environment and modifies the policy immediately resulting from the taken actions.

\subsubsection{MDP} The MDP offers a mathematical framework for simulating problem-solving situations where decisions are made by agents or decision-makers and where results are somewhat influenced by randomness. Studying optimization issues that can be addressed using dynamic programming and RL methods is made possible by MDPs. 
An MDP cam be described by the tuple $(S,A,P,R)$. 
$S$ is the finite set of predefined states, $A$ is the finite set of predefined actions the agent can choose in each state and $P$ is a transition probability matrix from state $s$ to $s^\prime$ resulting from action $a$. 
$R$ is a set of immediate rewards by taking each action. The mapping from a state to an action is 
determined by "policy" $\pi$. 
Finding the best policy to maximize the reward function is the purpose of an MDP. Its time horizon might be either finite or infinite. An ideal strategy $\pi^*$ that maximizes the anticipated total reward is defined for the MDP with a finite time horizon as

\begin{equation}
max_\pi E\left[\sum_{t=0}^{T} r_t(s_t,\pi(s_t))\right],
\label{eq:eq1}
\end{equation}
where $a_t = \pi(s_t)$. The goal for the infinite time horizon MDP can either be to maximize the average reward or the expected discounted total reward. The former is formulated as
\begin{equation}
    max_\pi E\left[\sum_{t=0}^{T}\gamma r_t(s_t,\pi(s_t))\right]
    \label{eq:eq2}
\end{equation}
and the latter as
\begin{equation}
    \lim inf_{T\to\infty}max_\pi E \left[\sum_{t=0}^{T} r_t(s_t,\pi(s_t))\right].
    \label{eq:eq3}
\end{equation}

In this research we are looking for analyzing the UAV trajectory from point $A$, which is our takeoff point, to point $B$, which is the landing point according to Fig. \ref{fig:gridworld}. While flying 
there different sub-regions in which the UAV is supposed to transmit the received data from the UEs to the CU/DU, as illustrated in Fig. \ref{fig:UAV movement}. Each sub-region has a set of IoT nodes transmitting data to the UAV at the same time which causes interference in the UAV receiver. 
We choose two reward functions:
\begin{equation}
    \begin{split}
        R_{PL} &= \frac{1}{1 + e^{-x}}\\
        R_{PL^{-1}} &= \frac{1}{1 + e^{-\frac{1}{x}}},
    \end{split}
    \label{eq:reward}
\end{equation}
where $x$ corresponds to the path loss defined in the system model.

\indent 
Two different scenarios presented for analysis: In one the UAV will follow the trajectory with the highest path loss since the reward will be higher and in the other, the UAV will follow a trajectory with lowest path loss which means being closer to the sub-region with higher received power. Trajectories with higher path-losses cause lower received power; hence the UAV should get closer to that sub-region. Moreover, taking an action would lead to a reward of $-1$ unless the UAV moves to either a new state or closer to the terminal state. 
The path-loss is applied to a sigmoid function which is bounded to $[0,1]$.\\

\begin{figure}
    \centering
    \includegraphics[scale = 0.4]{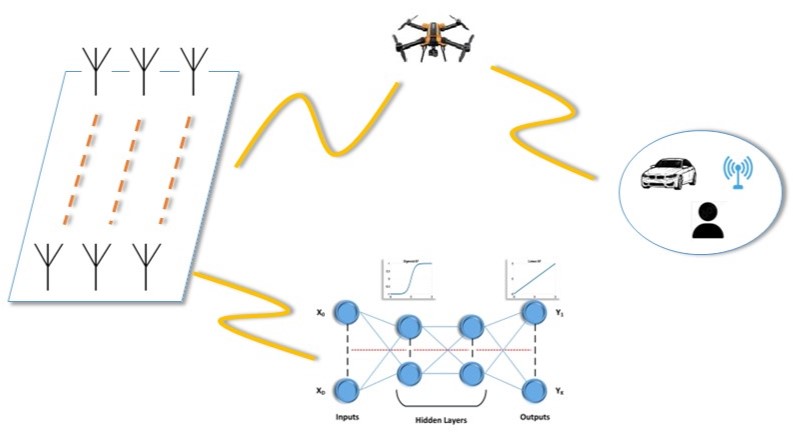}
    \caption{Uplink channel from UEs to UAV and downlink channel from UAV to CU with MLP }
    \label{fig:mimo-connection}
\end{figure}

\section{Simulation Results and Analysis}
\label{sec:analysis}
\begin{figure}
    \centering
    \includegraphics[scale=0.12]{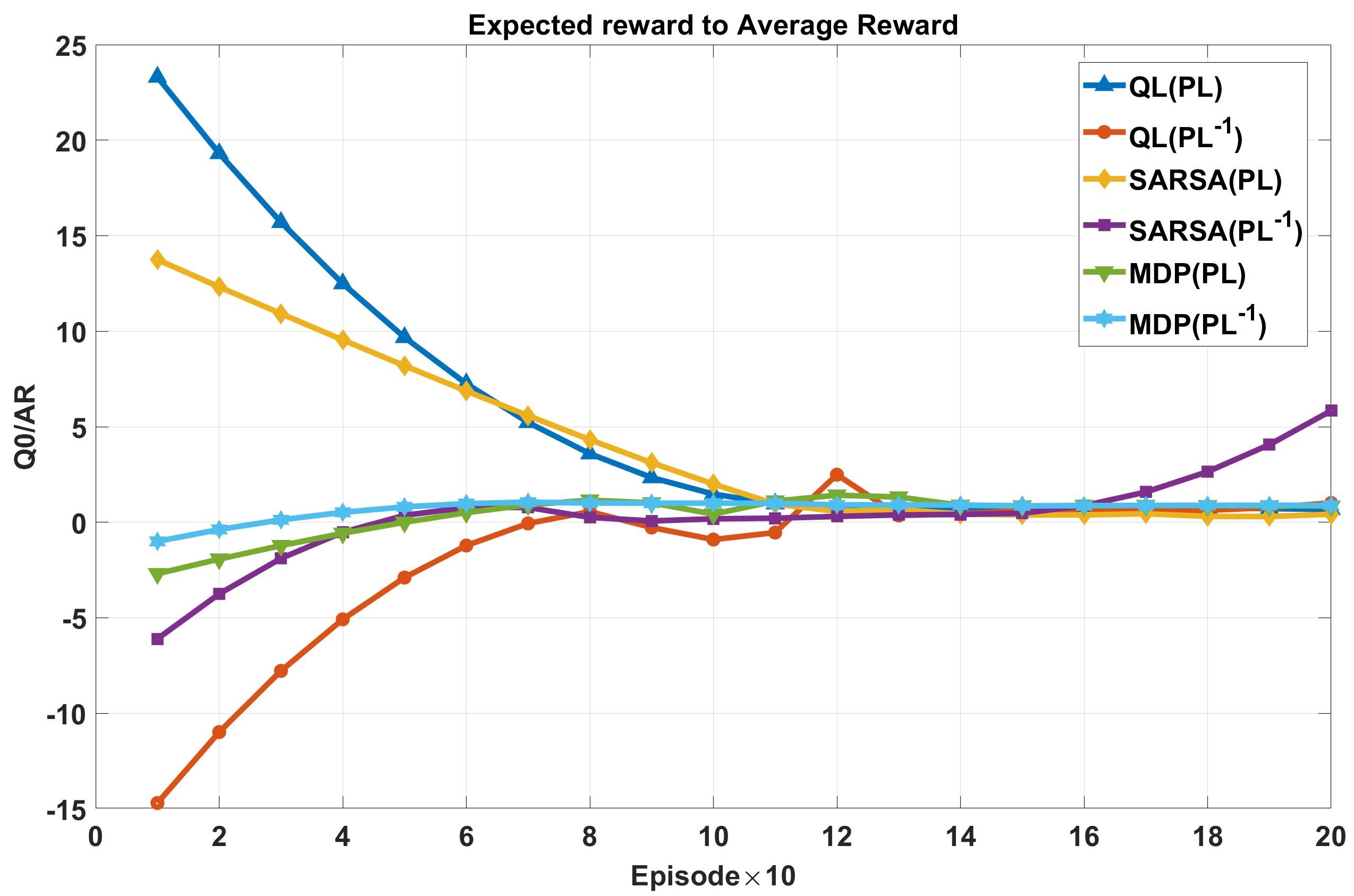}
    \caption{Expected reward at the beginning of each episode to the average reward}
    \label{fig:qtor}
\end{figure}

\begin{figure}
    \centering
    \includegraphics[scale=0.4]{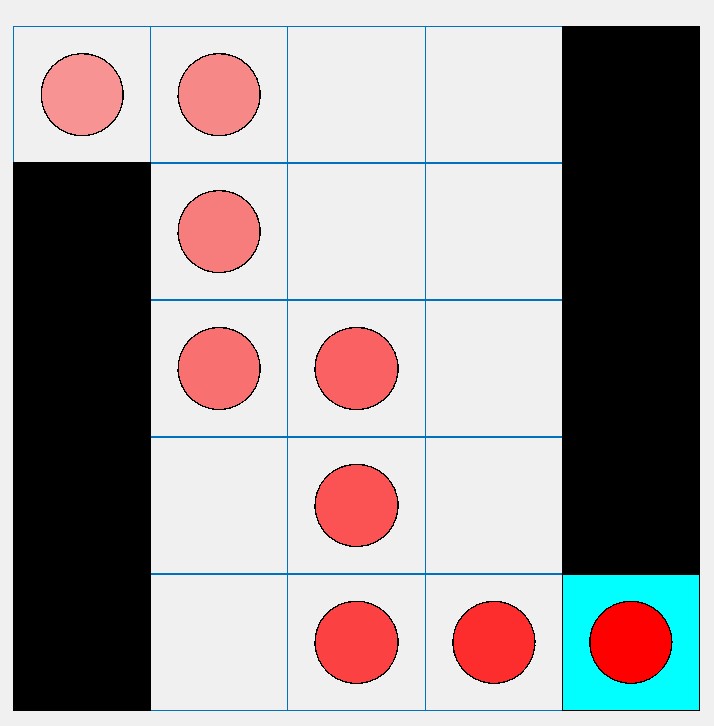}
    \caption{UAV trajectory from initial point to the terminal state}
    \label{fig:trajectory}
    \vspace{-3mm}
\end{figure}

\indent For the simulation, carrier frequency is $f_c = 6$GHz and the area which the UAV can move is $400 \times 400 m^2$. Each sub-region has an area equal to $80 \times 80$. Therefore, if we assume that each sub-region center is the center of cartesian plane, the UAV movement in each region has a uniform distribution between $[-40,40]$; otherwise, handover happens. Moreover, the altitude has the range of $[100,200]$ m. The path-loss is obtained from (\ref{eq:pl}) based on the distance between the UAV and the sub-region it is flying over. Table \ref{tbl:param} summarizes the simulation parameters.\\
\indent Flying at a consistent speed rather than hovering produces the most power-efficient operation $V=20m/s$ \cite{19bliss2020power},\cite{20zeng2019energy}. Therefore, for the access channel the receiver and for the fronthaul channel the transmitter are moving at the same speed. We assume that there are $2$ antennas in each sub-region for transmitting the data to the UAV and $4$ antennas at the UAV,  $2$ for receiving and $2$ for transmitting. Moreover, a rectangular array with $100$ antennas ($25 \times 4$) are employed at the CU/DU receiver.\\
\begin{table}
\centering
\caption{Simulation parameters}
\begin{tabular}{|c | c| c| c|} 
 \hline
 Parameter & Value & Parameter & Value\\ [0.5ex] 
 \hline\hline
 $\mu_{1,2}$ & 9.6 & $f_{c}$ & 6GHz \\ 
 \hline
 $\psi_{1,2}$ & 0.15 & $\alpha$ & 0.9 \\
 \hline
 $\eta_{LoS}$ & 1dB & $\gamma$ & 0.8 \\
 \hline
 $\eta_{NLoS}$ & 20dB & $x$ & [0 400] \\
 \hline
 $\Delta$ & 0.1 & $y$ & [0 400]\\
 \hline
 $\epsilon_{greedy}$ & 0.3 & $h$ & [100 200]\\
 \hline
 $n_{t_{UE}}$ & 2 & $n_{r_{UAV}}$ & 2\\
 \hline
 $n_{t_{UAV}}$ & 2 & $n_{r_{CU}}$ & 100\\ [1ex] 
 \hline
\end{tabular}
\vspace*{4mm}
\label{tbl:param}
\vspace{-7mm}
\end{table}
\begin{figure}
    \centering
    \includegraphics[scale=0.12]{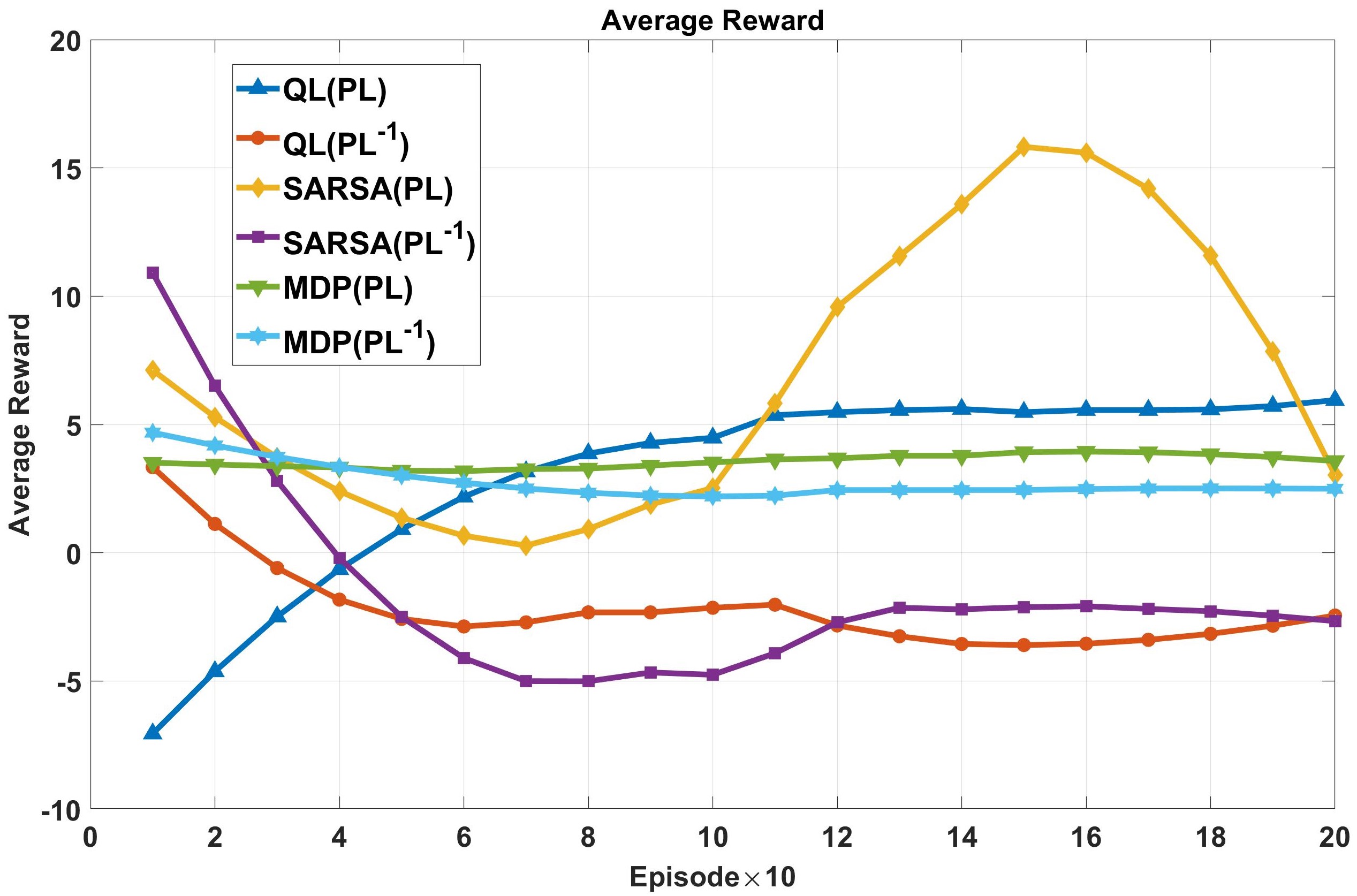}
    \caption{Average rewards for Q-learning, SARSA and MDP based on (\ref{eq:reward})}
    \label{fig:AR}
\end{figure}
\indent Fig. \ref{fig:AR} shows the average reward for different methods that the UAV can follow to get to the terminal state. Among Q-Learning, SARSA, and MDP, Q-Learning provides a smooth reward as the time episode increases for learning. It should be noted that this is for the case $PL$ and $R$ has the direct relation, which means that the UAV goes through the direction with the lower received power. In contrast, the average reward deteriorates if the UAV wants to follow the path of lower path-loss. For SARSA the result is a little different. Since SARSA is an online RL and updates itself in each iteration based on the environment and current state, we can get the maximum reward if the UAV has enough time to analyze the environment with the cost of more power consumption. Finally, for MDP the average reward has an almost constant shape which is appropriate for an unknown environment and states. 
Fig. \ref{fig:qtor} provides insights for finding the best discount factor and how much the UAV should care about the future reward to get to the terminal point. As the figure shows, MDP reaches to this condition ($\Delta \leq 10\%$) sooner at the cost of a lower average reward. However, SARSA is the slowest but has a higher average reward. Figure \ref{fig:trajectory} the final UAV trajectory from the takeoff point to the landing point for SARSA. It shows that the UAV chooses the shortest path to get to the terminal state. Further analysis of different modulation and MIMO fading channel has been left for future research.\\

\balance
\section{Conclusions}
\label{sec:conclusions}
In this study, we presented an aerial radio unit that connects to the distributed unit through a wireless fronthaul between the UAV and the nearest tower and dynamically serves an underserved area. We solved the UAV trajectory problem employing different RL solutions while the UAV needs to maintain simultaneously links to the UEs it serves and the CU/DU in a MIMO fading channel. In order to solve the problem of maximizing the network throughput based on UAV location, we first showed that it is nonconvex, and then we offered two different machine learning solutions. Assuming at first that the environment is a gridworld, we then allowed the UAV to explore the area by flying from point A to point B, using the offline Q-learning and the online SARSA algorithms, with the path-loss determining the reward. The trajectory in the second scenario is characterized as a Markov decision process with the goal of maximizing the average reward. Simulation results demonstrate that MDP achieves better results in a more constrained environment and with less effort. In contrast, SARSA requires a longer flight time but performs better in wider areas.


\bibliographystyle{ieeetr}
\bibliography{bib/main}

\end{document}